\documentclass[prl,%
 reprint,
 amsmath,amssymb,
 aps,
superscriptaddress]{revtex4-2}

\usepackage{graphicx}
\usepackage{dcolumn}
\usepackage{bm}
\usepackage{hyperref}

\usepackage{graphicx}
\usepackage{dcolumn}
\usepackage{bm}
\usepackage{bbm}
\usepackage{soul}
\usepackage{enumitem}
\usepackage{makecell}

\usepackage{braket}
\usepackage{graphicx}
\usepackage{amsmath,verbatim,latexsym,amssymb,indentfirst,mathrsfs,mathtools,amsthm,bbm,bm,hyperref,url}

\usepackage[title,titletoc]{appendix}

\usepackage{cancel}
\usepackage{bbold}
\usepackage[dvipsnames]{xcolor}
\usepackage[normalem]{ulem}
\usepackage{physics}

\begin{document}

\preprint{APS/123-QED}

\title{Real-time Monitoring of Neon Film Growth for Electron-on-Neon Qubits}

\author{Sidharth Duthaluru}
    \affiliation{Department of Physics, Washington University, Saint Louis, MO, USA, 63130.}
\author{Kaiwen Zheng}
    \affiliation{Department of Physics, Washington University, Saint Louis, MO, USA, 63130.}
\author{Erik A. Henriksen}
    \affiliation{Department of Physics, Washington University, Saint Louis, MO, USA, 63130.}
\author{Kater W. Murch}%
    \affiliation{Department of Physics, Washington University, Saint Louis, MO, USA, 63130.}
    \affiliation{Department of Electrical Engineering and Computer Science, University of California Berkeley, Berkeley, CA, USA, 94720.}
    \affiliation{Department of Physics, University of California Berkeley, Berkeley, CA, USA, 94720}

\date{\today}

\begin{abstract}
Electron-on-neon (eNe) charge states coupled to superconducting circuits are a promising platform for quantum computing. Control over the formation of these charge states requires techniques to track and control the growth of solid Ne films on the circuit surface. We demonstrate a real-time Ne film-growth monitor using high-transition-temperature (high-$T_c$) YBCO microwave resonators. The high $T_c$ enables tracking of the film thickness near Ne's triple temperature and below. Across more than 300 solidification experiments, we find that the final Ne thickness varies stochastically from a few nm to a few $\mu$m for films solidified from the liquid phase. By increasing the driving power in the resonator, we consistently reduce the final thickness to below 100 nm. These results represent an important step toward controlled formation of Ne films for eNe qubits and highlight the broader utility of high-$T_c$ resonators for hybrid quantum systems.

\end{abstract}

\maketitle

\section{Introduction}

Trapped single electrons on the surface of solid neon \cite{Cole_electron_trapping, Kajita_SN, Kajita_SN2}, coupled to a circuit quantum electrodynamics (circuit QED) architecture \cite{cQED_review}, constitute a promising emerging platform for quantum information processing \cite{Dafei_Ne}. Recent work has demonstrated impressive coherence times \cite{Dafei_NatPhys}, stable operation at elevated temperatures \cite{Xinhao_noise_resilient25}, and simultaneous coherent manipulation of multiple electrons \cite{Xinhao_interactingelectronqubits}. Despite such progress, the detailed mechanism underlying electron trapping remains poorly understood, with work identifying substrate \cite{Zheng2025} and surface roughness \cite{toshi,  Kanai_lateral_25} as key parameters in the formation of eNe charge states.  In addition, noble gas solids have long served as host media for a range of applications including 
inert trapping sites for unstable ions and molecules \cite{Nemukhin2007, Andrews1979, Tsegaw2021}, single atom detection substrates for low-yield nuclear reactions \cite{Loseth2019}, precision spectroscopy \cite{Lambo2021, Lancaster2024, Braggio2022},  and implantation hosts for solid state qubits \cite{Kanagin2013, Upadhyay2016, Kanagin25}. In all of these applications, the properties and crystal quality of the solid play a critical role.

There are two established ways to produce a solid noble gas film: quench condensing a gas directly into the solid phase on a cold substrate, and cooling a liquid film across the triple point into the solid phase. Prior work has investigated the resulting film qualities, revealing rough and porous solids from quench condensation \cite{Venables1968, Weiss1998, Metcalf2000}. Annealing techniques have been investigated to improve the quality of quench condensed films \cite{Kanagin2013}. From the liquid phase, prior work demonstrated that noble gas films exhibit triple point wetting \cite{triplePointWetting, LeidererReview}, where a liquid rapidly thins to vanishing thickness when slowly cooled past the triple temperature. However, recent work on eNe charge states \cite{Dafei_Ne, Zheng2025} has successfully produced Ne films from the liquid phase with thicknesses of tens of nanometers---substantially thicker than the few-atomic-layer limit predicted by triple point wetting. Motivated by the applications of noble gas films and the apparent challenge  of film growth, there has been renewed interest in film growth techniques \cite{LeidererReview, Cassidy_Ne_growth, Pop_hydrogen}.  In this work, we demonstrate a system to monitor film thickness dynamics in real time using a high-quality-factor YBCO microwave resonator \cite{YBCO_res1, YBCO_res2, YBCO_res3, YBCO_res4}. The high $T_{c}$~\cite{YBCO_discovery} of YBCO ensures that the resonator remains high-Q near the Ne triple temperature at 24.56~K~\cite{Ne_TP_value}. With this monitor, we study both liquid-phase and quench condensation growth techniques. We further characterize the effect of Ne volume, cooling rate, and microwave driving power on the dynamics of the Ne thickness in the vicinity of Ne's triple point. 

\section{Neon phase diagram and triple point}

\begin{figure}
\centering
    \includegraphics[width = \columnwidth]{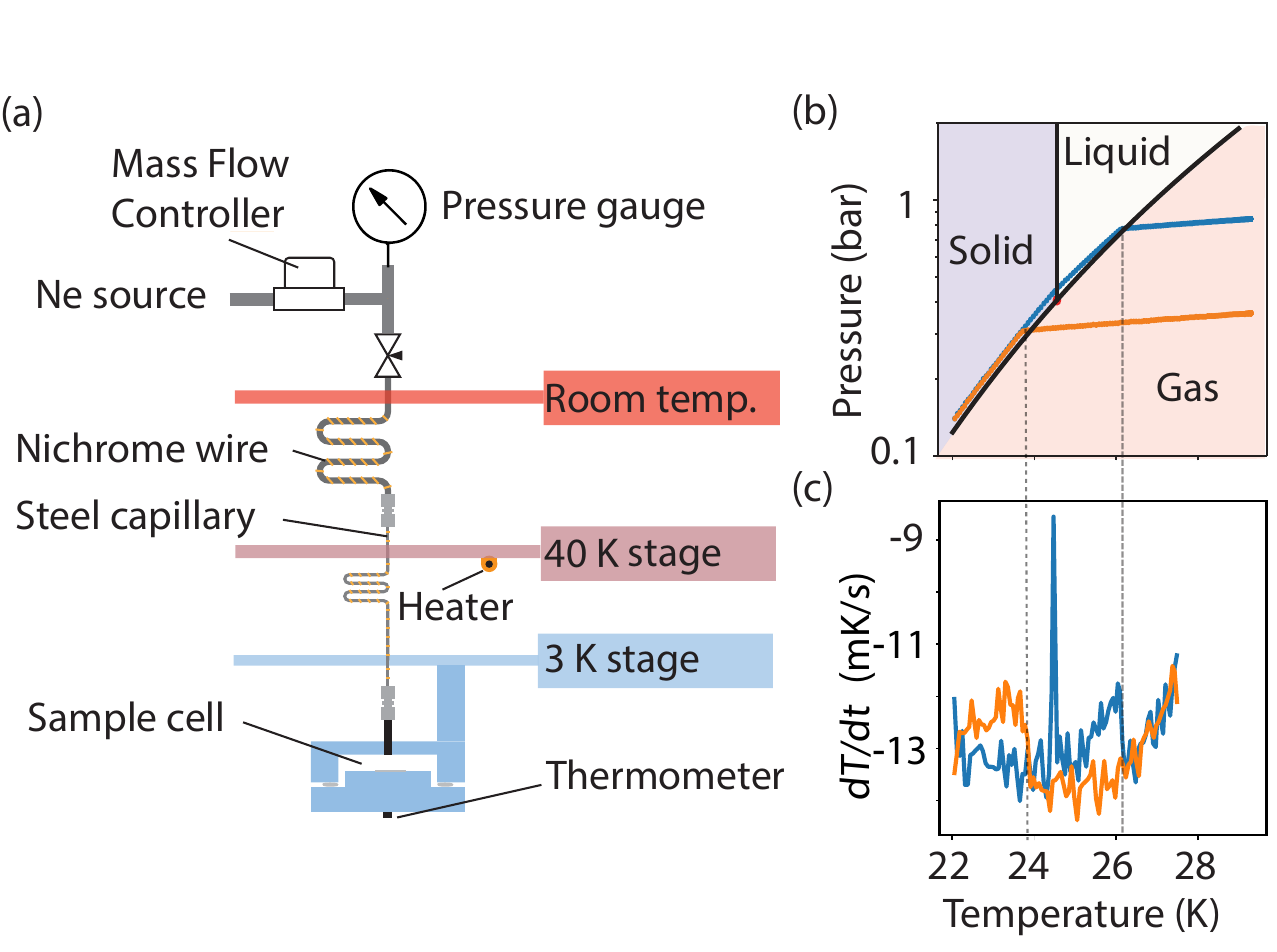}
    \caption{\textbf{Experimental setup to monitor cooling trajectories.} (a) Diagram of the experimental setup in a 3~K cryostat. A capillary from room temperature feeds into the sample cell, enabling real-time Ne deposition and pressure monitoring. The sample cell is anchored to the low-temperature (3~K) stage of the cryostat. The capillary is heated with a nichrome wire, and the temperature of the sample cell is controlled by applying heat to the intermediate (40~K) stage of the cryostat, ensuring that the sample cell is the coldest part of the gas system. (b) The Ne phase diagram, with two illustrated deposition trajectories crossing the triple point (red dot) at 24.56 K; gas$\to$liquid$\to$solid is shown in blue, while gas$\to$solid is shown in orange. (c) The sample cell temperature rate of change versus temperature shows the effects of latent heat at different phase transitions. The gray dashed lines demark the condensation and deposition temperatures for the two trajectories. The spike in $dT/dt$ is the result of the latent heat of freezing while crossing the triple temperature. }
    \label{fig1}
\end{figure}

We utilize a hermetically-sealed sample cell to produce specific trajectories on the pressure-temperature (P--T) phase diagram of Ne.  The sample cell is mounted to the 3~K stage of a variable-temperature cryostat (Fig.~\ref{fig1}(a)). A 1.59~mm inner diameter capillary connects the cell to a gas manifold at room temperature. We regulate Ne flow into the sample cell using a mass flow controller (MFC). A $\mathrm{RuO_x}$-based thermometer on the sample cell and a pressure gauge on the gas manifold yield the P--T trajectory of the sample cell. In Fig.~\ref{fig1}(b), we show two different trajectories along with the Ne phase diagram. The two trajectories correspond to different amounts of Ne gas injected at 29.5~K and cooled at to a final temperature of 22~K. For both trajectories, the initial pressure and temperature correspond to the gas phase. As the cell is cooled, the pressure decreases linearly in accordance with the ideal gas law. Once the Ne gas pressure and temperature reach the phase boundary, portions of the gas start to condense. Depending on whether the temperature is above or below the triple temperature, the condensed phase is liquid or solid. As shown by the blue trajectory in Fig.~\ref{fig1}(b), $0.008$~mol of injected gas condenses into a liquid at $\sim$26~K. Further cooling follows the phase boundary and crosses the triple point into the solid phase. Figure~\ref{fig1}(c) displays small changes in the observed sample cell cooling rate corresponding to the latent heat of condensation and freezing.  By contrast, when injecting $0.003$ mol (orange curve), the system reaches the saturation pressure at 23.8~K, and the film starts to condense into the solid phase below the triple temperature. The latent heat of deposition can be seen in the cooling rate change.

Based on this phase diagram, one would expect the final Ne film thickness to depend on the volume of Ne initially injected. Thin films would be obtained from a small amount of injected gas and cooled directly into the solid portion of the phase diagram. Thicker films would require a trajectory that passes through the liquid phase and into the solid phase. Contrary to this intuition, prior work identified that cooling trajectories through the triple point produced vanishingly thin films---a phenomenon referred to as triple point wetting \cite{triplePointWetting, LeidererReview}, where the film thins to a few atomic layers when cooled past the triple point.  Such thin films may be unsuitable for eNe qubits as electrons can tunnel through the Ne film into the underlying substrate. However, recent experiments have demonstrated few to tens of nm thick liquid-phase-grown films for trapping electrons \cite{Dafei_Ne,Zheng2025}, in disagreement with the predictions from triple point wetting \cite{triplePointWetting, LeidererReview}. To resolve the discrepancy, a detailed understanding of film thickness and morphology for different growth methods is desired. 

\section{High-$T_c$ YBCO Resonator thickness MoNiToR}

\begin{figure*}
\centering
    \includegraphics[width = \textwidth]{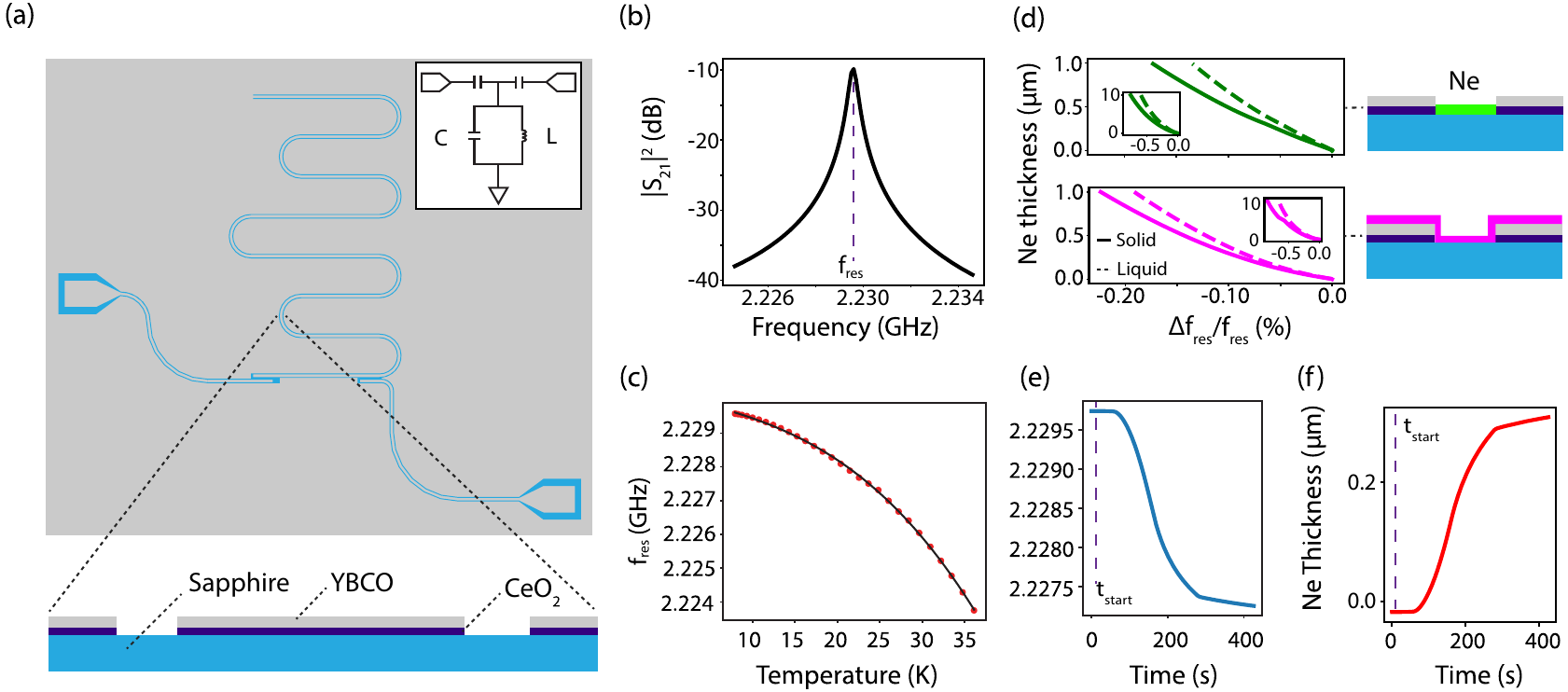}
    \caption{\textbf{YBCO resonator thickness monitor.} (a) Resonator design. The lower panel shows a (not to scale) cross-section of the chip with YBCO grown on a CeO$_2$ seed layer. The inset shows the corresponding lumped-element circuit diagram of a capacitor and an inductor in parallel to ground. (b) The resonator transmission $|S_{21}|^2$ versus frequency.  (c) Measured  $f_\mathrm{res}$ versus temperature. (d) Finite element modeling is used to estimate the Ne thickness from fractional frequency shifts of the resonator. Magenta is the conformal model, while green is the in-trench model. The dashed and solid lines correspond to liquid and solid phases, respectively. Side panels illustrate both Ne models on the chip surface. (e) Real-time tracking of Ne deposition onto the resonator at low temperature. At $t_\mathrm{start}$, Ne is injected via the gas manifold and $f_\mathrm{res}$ is recorded versus time. (f) Solid Ne thickness tracking over time, assuming a conformal growth morphology.}
    \label{fig2}
\end{figure*}

We incorporate a high-$T_c$ YBCO microwave resonator into our hermetic sample cell to monitor Ne film thickness dynamics in real time.  When Ne is deposited on the surface of the resonator, its relative permittivity ($\epsilon_\mathrm{solid} = 1.22$ for the solid phase, and $\epsilon_\mathrm{liquid} = 1.19$ for the liquid phase) alters the characteristic capacitance of the resonator, and shifts its frequency. Continuous frequency tracking of the resonator can therefore be used to infer the Ne film thickness in real time. 

The 200-nm thick YBCO film is deposited epitaxially on r-plane sapphire with a 40~nm CeO$_2$ buffer layer for lattice matching \cite{trueYBCO_epi, CeO2_buffer}. The film is then patterned using standard photolithography and dry-etched using an Ar ion milling process to form microwave frequency resonators \cite{filmnote}. 
The resonator we use in this study is a $\lambda/4$ coplanar waveguide resonator (CPW) resonator that is capacitively coupled to input and output CPW feedlines (See Fig.~\ref{fig2}a) \cite{CPW_cQED, McRae_review}. The simplified equivalent circuit representation as a lumped element LC resonator is shown in the inset.   The CPW resonator has a central conductor linewidth of 30~$\mu$m and a gap of 10~$\mu$m to the ground plane.  The fabrication process produces an 80~nm overetch into the sapphire substrate, forming a 320~nm-deep trench.  We measure the scattering matrix element, $S_{21}$, of the resonator using a vector network analyzer (VNA).  Figure~\ref{fig2}(b) is the resulting transmission response.  At $T=5$~K the resonator has a loaded quality factor of 6200, and a resonant frequency $f_\mathrm{res}$ of $\sim2.230$~GHz.

While the experiments performed here are conducted well below the $T_c$ of YBCO, remnant equilibrium quasiparticle population \cite{YBCO_lambda, Hirschfeld_YBCO_1, Hirschfeld_YBCO_2, YBCO_res1, YBCO_res2, YBCO_res3, YBCO_res4} and two-level system saturation \cite{TLS_original, TLS_freq_shift, muller_TLS} result in a temperature dependence of the resonator's quality factor and frequency. To isolate the capacitive shift from Ne alone, we calibrate the monitor across temperatures from 5 to 35~K and establish a reference baseline (Fig.~\ref{fig2}(c)). We fit this temperature dependence to a 3rd-order polynomial function. This allows us to apply a temperature-dependent frequency correction to all measured resonance data across the full temperature range.

When Ne accumulates on the surface of the CPW resonator, the increased dielectric constant alters the characteristic capacitance of the waveguide, changing its resonant frequency \cite{GeYang, Koolstra, Dafei_Ne, Zheng2025, Cassidy_Ne_growth}. For Ne films deposited from the gas phase, we expect it to form a conformal coating on the sample surface. For Ne films grown from the liquid phase, we apply a different model where Ne fills the trench region first.  Using these models and finite element simulation, we can infer the effective thickness of Ne films based on the fractional frequency shift $\Delta f_\mathrm{res}/f_\mathrm{res}$, as shown in Fig.~\ref{fig2}(d).

To demonstrate real-time film thickness monitoring, we track the resonator frequency over time as a small volume of Ne gas is injected into the sample cell at $T=8$~K, as shown in Fig.~\ref{fig2}(e). Since the vapor pressure at this temperature is vanishingly small, this injection of gas corresponds to a non-equilibrium quench condensation method, where Ne accumulates on the cold sample surface as a solid.   At $t_\mathrm{start}=0$, $1.78\times10^{-4}$~mol of Ne is released into the sample cell at a constant rate of 1 sccm via the gas manifold. We observe a subsequent decrease in the resonator frequency as the gas travels down the capillary tube and condenses onto the resonator surface. Based on the thickness calibration, the frequency shift can be converted to a Ne thickness versus time as displayed in Fig.~\ref{fig2}(f).

\section{Real-time monitoring of solid neon growth}

\begin{figure}
    \centering
    \includegraphics[width = \columnwidth]{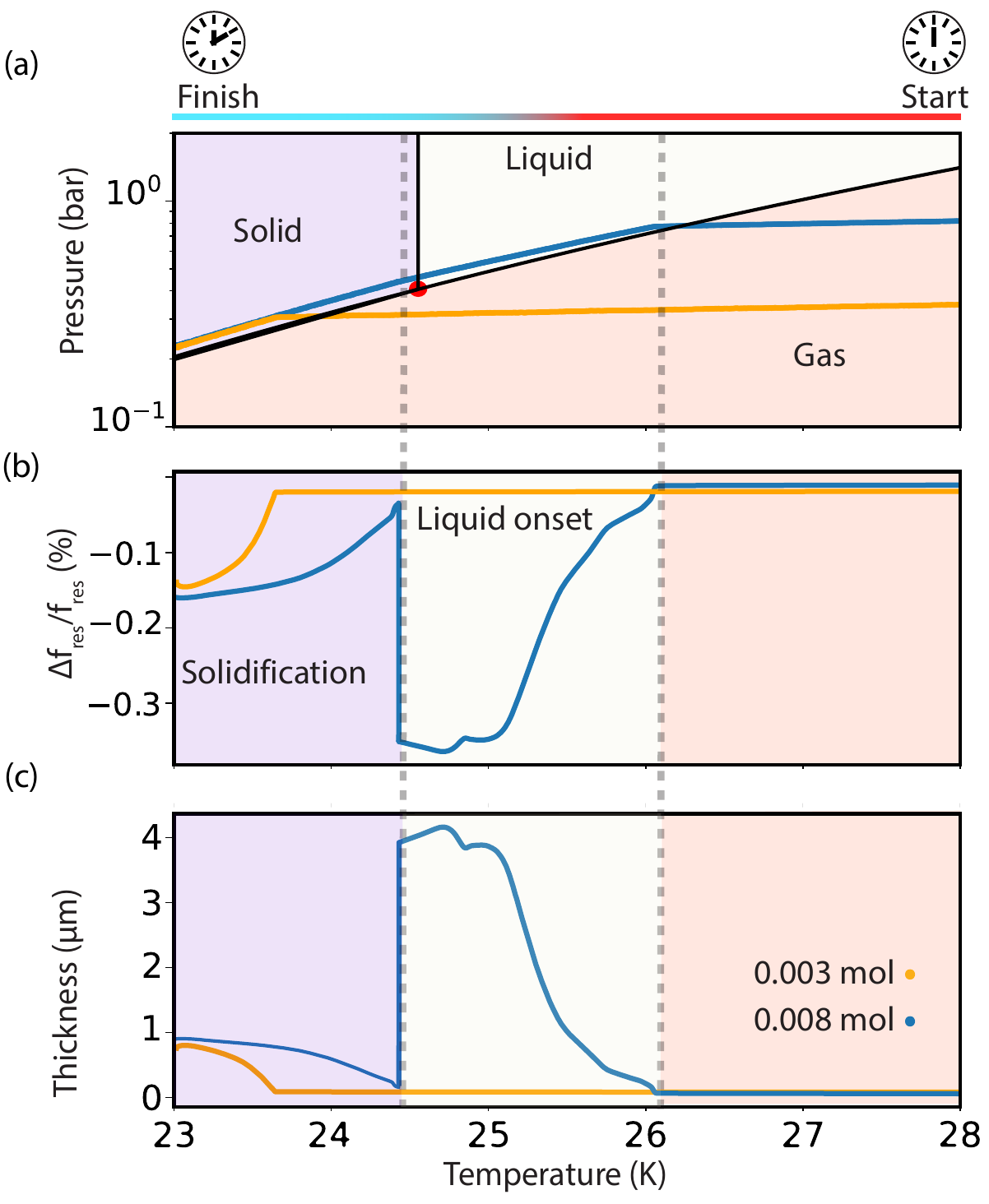}
    \caption{\textbf{Liquid and solid film growth monitoring.} (a) The Ne phase diagram, (b) observed resonator frequency shifts, and (c) inferred film thickness versus temperature.  Two initial gas quantities: 0.008 mol (blue) and 0.003 mol (orange) correspond to gas$\to$liquid$\to$solid and gas$\to$solid deposition trajectories respectively. The sample cell cooling rate is 0.07 K/min. The gray dashed lines denote approximate temperature regions of liquid film growth and solidification for the blue trajectory.}
    \label{fig3}
\end{figure}

Having demonstrated the function of the microwave-resonator-based thickness monitor, we return to the two quasi-equilibrium growth scenarios presented in Fig.~\ref{fig1}(b).  Figure~\ref{fig3} displays the phase diagram trajectories, observed resonator frequency shifts, and inferred thickness for the gas$\to$solid (orange), and gas$\to$liquid$\to$solid (blue) deposition trajectories.  For the gas$\to$solid trajectory, the film thickness starts to increase at the temperature where the Ne reaches the vapor--solid coexistence line.  Here, the fraction of the Ne that remains in the gas phase is set by the equilibrium condition at the saturation pressure; lower temperatures correspond to lower saturation pressures and a larger fraction in the solid phase.

For the gas$\to$liquid$\to$solid deposition trajectory, the system reaches the vapor--liquid coexistence line above the triple temperature, and the equilibrium condition results in an accumulation of Ne in the liquid phase. This can be seen as a significant increase in the thickness of the liquid film at temperatures below 26~K.  As the sample cell crosses the triple temperature, we then observe a rapid thinning of the film, leaving a film of vanishing thickness (blue line) \footnote{We attribute the difference in solidification temperature from the nominal triple temperature (24.56 K) to a slight miscalibration of the thermometer.}.  This observation is in agreement with prior studies of triple point wetting \cite{triplePointWetting, LeidererReview} where the substrate--adsorbate interaction dominates over the adsorbate--adsorbate interaction at the triple temperature, resulting in a transition from bulk liquid to a thin conformal solid coating the entire sample cell.  As the sample is further cooled, we observe an increasing solid film thickness, which corresponds to the gas$\to$solid phase transition with excess gas adhering to the thin Ne coating.  Notably, despite starting with a 2.7 times larger volume of Ne with gas$\to$liquid$\to$solid deposition trajectory, we observe only a 6\% increase of eventual solid film thickness compared to the gas$\to$solid trajectory.

\section{Controlling neon solidification dynamics across the triple point}

\begin{figure}
    \centering
    \includegraphics[width = \columnwidth]{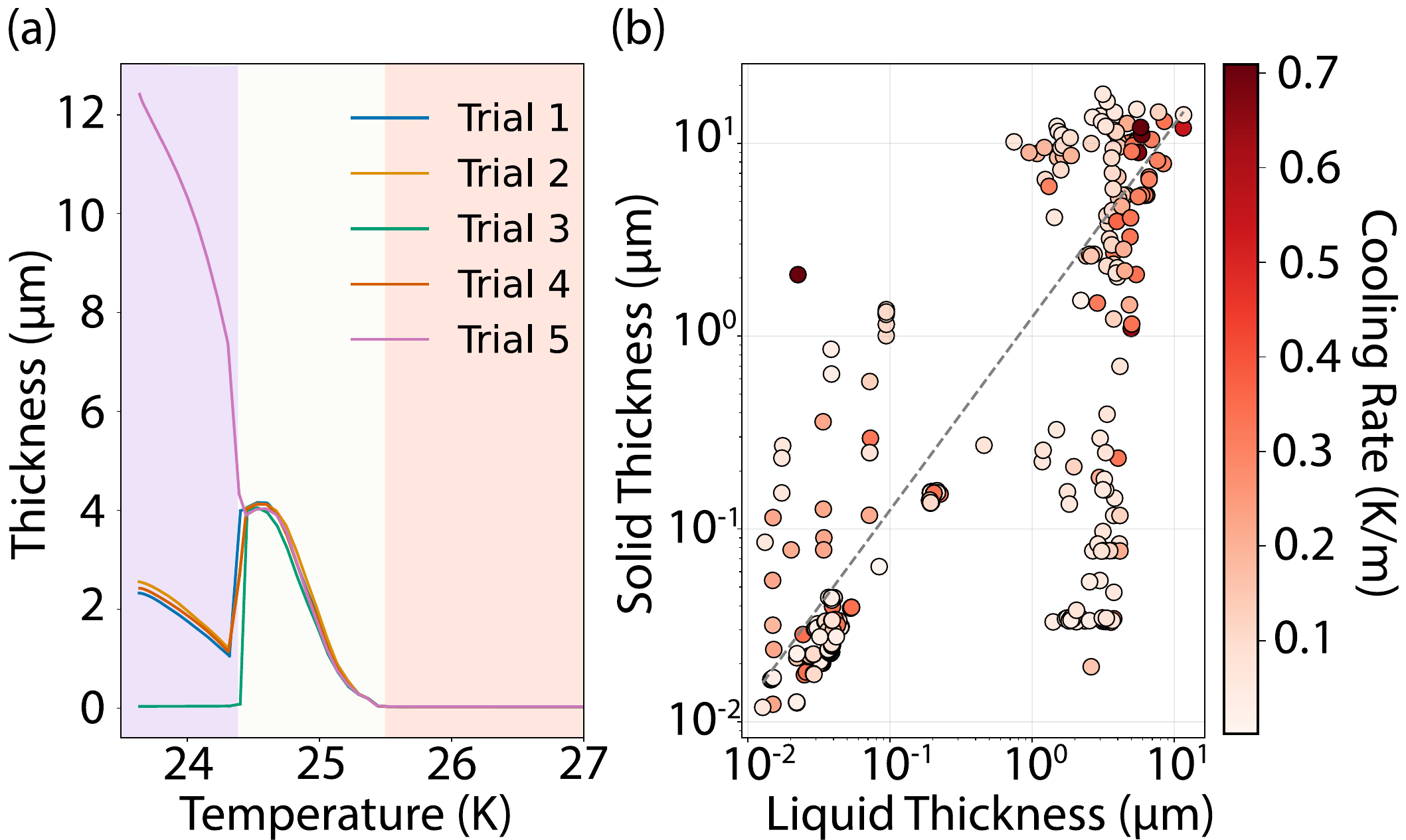}
    \caption{\textbf{Stochastic thickness of solid Ne films.} (a) Five consecutive gas$\to$liquid$\to$solid deposition trajectories (0.006~mol Ne). For all trajectories, the onset of liquid condensation occurs at 25.5~K. The final film thickness varies significantly for the repeated trajectories. (b) Comparison of the solid film thickness (at 23.8~K) to the liquid film thickness (at 24.7~K) across different cooling rates. The resulting solid film varies between 10~nm and several tens of $\mu$m with weak dependence on the initial liquid thickness (correlation factor $r=0.6$, gray dashed line). }
    \label{fig4}
\end{figure}

We now turn to a systematic characterization of solid Ne film thickness versus Ne volume and cooling rate.  Figure~\ref{fig4}(a) displays the inferred Ne thickness for five consecutive solidification trajectories with a cooling rate of 0.07~K/min, where the sample cell is repeatedly ramped between $T=29$ and 23.8~K.  For all five trials, we observe repeatable liquid growth to a thickness of $\sim$~4~$\mu$m. Below the triple temperature, the repetitions show varying trajectories: in three cases, the film reaches a final solid thickness of $\sim$~2.5~$\mu$m, while in one case the film is vanishingly thin and in another it reaches a thickness greater than 12~$\mu$m. This is our primary observation; the final thickness of the solid film exhibits significant stochastic variations even for the cases where rapid thinning at the triple temperature~\cite{triplePointWetting, LeidererReview} is observed.

To further explore this stochasticity, we perform a series of solidification experiments with different injected Ne volumes. After loading a fixed amount of Ne into the sample cell, we repeatedly cycle the sample temperature between 29~K and 23.8~K to induce melting and solidification while monitoring the film thickness. We test different cooling rates ranging between 0.03~K/min and 0.7~K/min. For each rate, we allow the Ne to fully melt and solidify several times. We repeat this procedure for  different injected Ne volumes to assess how the stochastic behavior depends on the amount of Ne present. We plot the solid Ne thickness at 23.8~K vs.\ the liquid Ne thickness at 24.7~K for a total of 364 such solidification events in Fig.~\ref{fig4}(b).  Although the solid film thickness is correlated with the liquid thickness (Pearson's correlation coefficient $r=0.6$), the final solid thickness grown from the liquid phase remains  unpredictable, spanning nearly three orders of magnitude for thick liquid films and two orders of magnitude for thin liquid films.

\begin{figure*}
    \centering
    \includegraphics[width = .9\textwidth]{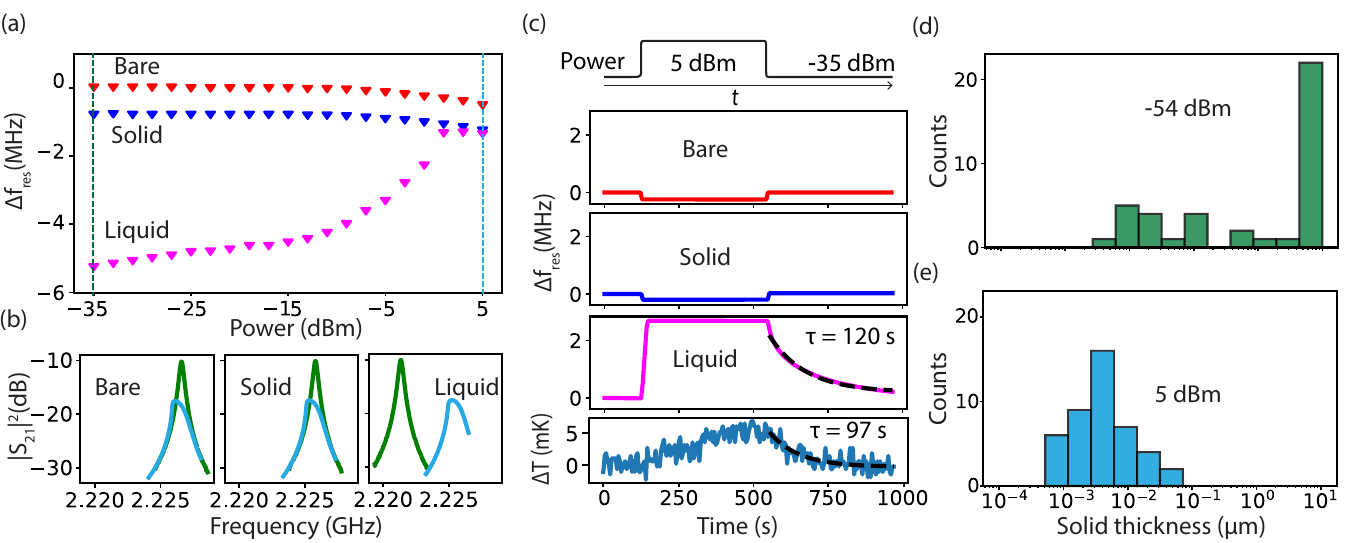}
    \caption{\textbf{Dependence of film dynamics on microwave driving power.} (a) The observed resonance frequency shift as a function of driving power in the bare resonator, solid Ne film (thickness $\approx$ 300~nm, 23.8~K), and liquid film (thickness $\approx$ 1$~\mu$m, 24.7~K). (b) Resonator frequency shift for all three cases from (a) with green and blue traces corresponding to $-35$~dBm and 5~dBm respectively.  (c) Abrupt reduction of the probe power results in near instantaneous changes in the profile for the bare and solid-Ne-covered resonator case. By contrast, we observe a slow relaxation of the frequency shift to its low power value for the liquid film. This behavior is accompanied by a relaxation of the sample cell temperature at a similar timescale. (d,e) Histograms of the final solid film thicknesses (at 23.8~K) when cooling with a $-54$~dBm and 5~dBm driving power.}   
    \label{fig5}
\end{figure*}

The unpredictable nature of film solidification motivates a search for control parameters that might stabilize the growth process. We therefore examine how microwave excitation of the resonator affects Ne film trajectories across the triple point. Figure~\ref{fig5}(a) displays the frequency shift from its baseline value as a function of driving power for three film conditions: no Ne (bare resonator) at 24.7~K, solid Ne at 23.8~K, and liquid Ne at 24.7~K. For the bare resonator case, the resonator frequency decreases with increasing driving power due to the Kerr nonlinearity~\cite{cQED_review}. As shown in Fig.~\ref{fig5}(b), the resonator line shape becomes asymmetric, approaching a bifurcating behavior \cite{JBA} at driving powers above $\sim- 5$~dBm. The same behavior is observed for the resonator with solid Ne and liquid Ne.  In the liquid Ne case, however, we observe an additional overall increase in resonator frequency when driven at higher powers. This is indicative of thinning of the liquid Ne. While prior work with micron-thick helium films observed that a strong direct-current (DC) bias can push the liquid out of a micro-channel~\cite{GeYang}, such an electrostriction effect at the GHz frequency excitation enlisted here is unlikely. 

Figure~\ref{fig5}(c) presents the time-domain frequency response of the resonator when the driving power increases from $-35$~dBm to 5~dBm and back to $-35$~dBm after a 7-minute delay. We observe that the frequency of the liquid-Ne-covered resonator gradually returns to the initial value with a time constant $\tau=120\pm3$~s. The accompanying change of the sample-cell temperature, $\Delta T$, exhibits a similar time constant of $97\pm19$~s. This suggests that the dramatic change in the Ne thickness for the liquid-Ne-covered case is likely due to localized heating caused by quasiparticle-related dissipation at high circulation powers.

The observed power dependence could be a valuable control knob for Ne film growth across the triple point. We demonstrate this by conducting identical sets of measurements at two different driving powers. For a given injected Ne volume, we test three cooling rates, and for each rate we record five melt--solidify cycles, yielding 15 solidification events at $-54$~dBm. We repeat the same procedure at 5~dBm using the same Ne loading for comparison. After completing measurements at both driving powers, we evacuate and reload the cell with a different Ne volume and repeat the full sequence. In total, we examine three Ne volumes, resulting in 45 solidification events at each power. For low driving power (Fig.~\ref{fig5}(d)), the thickness of Ne after solidification ranges from a few $\mu$m to a few nm. This is similar to the stochastic behavior shown in Fig.~\ref{fig4}(b), where the resonator is monitored at driving powers between $-25$ to $-35$~dBm. In comparison, for higher driving power (Fig.~\ref{fig5}(e)) the films are consistently thinner than 100~nm. 

\section{Outlook}

We have demonstrated a real-time Ne film growth monitor using a high-$T_c$ YBCO microwave resonator, enabling continuous tracking of film thickness dynamics from 5 to 35~K. Across more than 300 solidification experiments, we observed that Ne films grown through the liquid phase exhibit highly stochastic final thicknesses.  Importantly, we discovered that microwave driving power provides an effective control parameter: high probe powers  consistently produce films below 100~nm thickness, while low powers yield the full stochastic distribution. We attribute this control mechanism to localized heating at the resonator location, which redistributes liquid Ne within the sample cell before solidification.

Our observations reconcile perceived contradictions between triple-point wetting theory~\cite{triplePointWetting, LeidererReview} and recent eNe experiments~\cite{Dafei_Ne, Zheng2025}. While we observe the predicted film thinning at the triple temperature, subsequent gas deposition onto the thin wetting layer can produce films of 10--50~nm thickness at temperatures well below the triple point---consistent with the film thicknesses successfully used in current eNe qubit implementations at millikelvin temperatures.

Several avenues could extend this work toward improved control of Ne film growth. Spatial mapping of film distribution could be achieved through arrays of resonators along a transmission line \cite{Cassidy_Ne_growth} or by incorporating a scanning probe. Such spatial resolution, combined with optical characterization techniques, would reveal the relationship between film morphology, surface roughness, and electron trapping sites critical for eNe qubits. Additionally, systematic variation of the resonator cross-sectional geometry could provide further insights into the mechanisms governing solidification dynamics. These advances would enable deterministic control over Ne film properties essential for scalable electron-on-neon quantum devices and broader applications of noble gas substrates in quantum technologies.

\begin{acknowledgments}
The authors thank Paul Leiderer, Toshiaki Kanai, and Johannes Pollanen for fruitful discussions. The authors acknowledge Matthew K. Fowlerfinn and Yinyao Shi for their assistance with the experimental setup. This work is supported by the Gordon and Betty Moore Foundation, DOI 10.37807/gbmf11557 and the National Science Foundation award No.~2427093. The authors acknowledge the use of facilities at the Institute of Materials Science and Engineering in Washington University.

\end{acknowledgments}


\begin{thebibliography}{10}
\expandafter\ifx\csname url\endcsname\relax
  \def\url#1{\texttt{#1}}\fi
\expandafter\ifx\csname urlprefix\endcsname\relax\def\urlprefix{URL }\fi
\providecommand{\bibinfo}[2]{#2}
\providecommand{\eprint}[2][]{\url{#2}}

\bibitem{Cole_electron_trapping}
\bibinfo{author}{Cole, M.~W.} \& \bibinfo{author}{Cohen, M.~H.}
\newblock Image-Potential-Induced Surface Bands in Insulators.
\newblock \emph{\bibinfo{journal}{Phys. Rev. Lett.}}
  \textbf{\bibinfo{volume}{23}}, \bibinfo{pages}{1238--1241}
  (\bibinfo{year}{1969}).

\bibitem{Kajita_SN}
\bibinfo{author}{Kajita, K.}
\newblock A new two-dimensional electron system on the surface of solid neon.
\newblock \emph{\bibinfo{journal}{Surface Science}}
  \textbf{\bibinfo{volume}{142}}, \bibinfo{pages}{86--95}
  (\bibinfo{year}{1984}).

\bibitem{Kajita_SN2}
\bibinfo{author}{Kajita, K.}
\newblock Electronic Properties of Two Dimensional Electrons Formed on Liquid
  Helium and Solid Neon.
\newblock \emph{\bibinfo{journal}{Japanese Journal of Applied Physics}}
  \textbf{\bibinfo{volume}{26}}, \bibinfo{pages}{1943} (\bibinfo{year}{1987}).

\bibitem{cQED_review}
\bibinfo{author}{Blais, A.}, \bibinfo{author}{Grimsmo, A.~L.},
  \bibinfo{author}{Girvin, S.~M.} \& \bibinfo{author}{Wallraff, A.}
\newblock Circuit quantum electrodynamics.
\newblock \emph{\bibinfo{journal}{Rev. Mod. Phys.}}
  \textbf{\bibinfo{volume}{93}}, \bibinfo{pages}{025005}
  (\bibinfo{year}{2021}).

\bibitem{Dafei_Ne}
\bibinfo{author}{Zhou, X.} \emph{et~al.}
\newblock Single electrons on solid neon as a solid-state qubit platform.
\newblock \emph{\bibinfo{journal}{Nature}} \textbf{\bibinfo{volume}{605}},
  \bibinfo{pages}{46--50} (\bibinfo{year}{2022}).

\bibitem{Dafei_NatPhys}
\bibinfo{author}{Zhou, X.} \emph{et~al.}
\newblock Electron charge qubit with 0.1{\thinspace}millisecond coherence time.
\newblock \emph{\bibinfo{journal}{Nature Physics}}
  \textbf{\bibinfo{volume}{20}}, \bibinfo{pages}{116--122}
  (\bibinfo{year}{2024}).

\bibitem{Xinhao_noise_resilient25}
\bibinfo{author}{Li, X.} \emph{et~al.}
\newblock Noise-resilient solid host for electron qubits above 100 mK
  (\bibinfo{year}{2025}).
\newblock \bibinfo{note}{ArXiv:2502.01005}.

\bibitem{Xinhao_interactingelectronqubits}
\bibinfo{author}{Li, X.} \emph{et~al.}
\newblock Coherent manipulation of interacting electron qubits on solid neon
  (\bibinfo{year}{2025}).

\bibitem{Zheng2025}
\bibinfo{author}{Zheng, K.}, \bibinfo{author}{Song, X.} \&
  \bibinfo{author}{Murch, K.~W.}
\newblock Surface-Morphology-Assisted Trapping of Strongly Coupled
  Electron-on-Neon Charge States.
\newblock \emph{\bibinfo{journal}{Physical Review Letters}}
  \textbf{\bibinfo{volume}{135}} (\bibinfo{year}{2025}).

\bibitem{toshi}
\bibinfo{author}{Kanai, T.}, \bibinfo{author}{Jin, D.} \& \bibinfo{author}{Guo,
  W.}
\newblock Single-Electron Qubits Based on Quantum Ring States on Solid Neon
  Surface.
\newblock \emph{\bibinfo{journal}{Phys. Rev. Lett.}}
  \textbf{\bibinfo{volume}{132}}, \bibinfo{pages}{250603}
  (\bibinfo{year}{2024}).

\bibitem{Kanai_lateral_25}
\bibinfo{author}{Kanai, T.} \& \bibinfo{author}{Zhang, C.}
\newblock Electron Lateral Trapping Induced by Non-Uniform Thickness in Solid
  Neon Layers (\bibinfo{year}{2025}).
\newblock \bibinfo{note}{ArXiv:2510.10351}.

\bibitem{Nemukhin2007}
\bibinfo{author}{Nemukhin, A.~V.}, \bibinfo{author}{Khriachtchev, L.~Y.},
  \bibinfo{author}{Grigorenko, B.~L.}, \bibinfo{author}{Bochenkova, A.~V.} \&
  \bibinfo{author}{R\"{a}s\"{a}nen, M.}
\newblock Investigation of matrix-isolated species: spectroscopy and molecular
  modelling.
\newblock \emph{\bibinfo{journal}{Russian Chemical Reviews}}
  \textbf{\bibinfo{volume}{76}}, \bibinfo{pages}{1085--1092}
  (\bibinfo{year}{2007}).

\bibitem{Andrews1979}
\bibinfo{author}{Andrews, L.}
\newblock Spectroscopy of Molecular Ions in Noble Gas Matrices.
\newblock \emph{\bibinfo{journal}{Annual Review of Physical Chemistry}}
  \textbf{\bibinfo{volume}{30}}, \bibinfo{pages}{79--101}
  (\bibinfo{year}{1979}).

\bibitem{Tsegaw2021}
\bibinfo{author}{Tsegaw, Y.~A.} \emph{et~al.}
\newblock (Noble Gas)n-NC+ Molecular Ions in Noble Gas Matrices: Matrix
  Infrared Spectra and Electronic Structure Calculations.
\newblock \emph{\bibinfo{journal}{Chemistry --A European Journal}}
  \textbf{\bibinfo{volume}{28}} (\bibinfo{year}{2021}).

\bibitem{Loseth2019}
\bibinfo{author}{Loseth, B.} \emph{et~al.}
\newblock Detection of atomic nuclear reaction products via optical imaging.
\newblock \emph{\bibinfo{journal}{Physical Review C}}
  \textbf{\bibinfo{volume}{99}} (\bibinfo{year}{2019}).

\bibitem{Lambo2021}
\bibinfo{author}{Lambo, R.} \emph{et~al.}
\newblock High-resolution spectroscopy of neutral Yb atoms in a solid Ne
  matrix.
\newblock \emph{\bibinfo{journal}{Physical Review A}}
  \textbf{\bibinfo{volume}{104}} (\bibinfo{year}{2021}).

\bibitem{Lancaster2024}
\bibinfo{author}{Lancaster, D.~M.}, \bibinfo{author}{Dargyte, U.} \&
  \bibinfo{author}{Weinstein, J.~D.}
\newblock Optical spin readout of single rubidium atoms trapped in solid neon.
\newblock \emph{\bibinfo{journal}{Physical Review Research}}
  \textbf{\bibinfo{volume}{6}} (\bibinfo{year}{2024}).

\bibitem{Braggio2022}
\bibinfo{author}{Braggio, C.} \emph{et~al.}
\newblock Spectroscopy of Alkali Atoms in Solid Matrices of Rare Gases:
  Experimental Results and Theoretical Analysis.
\newblock \emph{\bibinfo{journal}{Applied Sciences}}
  \textbf{\bibinfo{volume}{12}}, \bibinfo{pages}{6492} (\bibinfo{year}{2022}).

\bibitem{Kanagin2013}
\bibinfo{author}{Kanagin, A.~N.}, \bibinfo{author}{Regmi, S.~K.},
  \bibinfo{author}{Pathak, P.} \& \bibinfo{author}{Weinstein, J.~D.}
\newblock Optical pumping of rubidium atoms frozen in solid argon.
\newblock \emph{\bibinfo{journal}{Physical Review A}}
  \textbf{\bibinfo{volume}{88}} (\bibinfo{year}{2013}).

\bibitem{Upadhyay2016}
\bibinfo{author}{Upadhyay, S.} \emph{et~al.}
\newblock Longitudinal Spin Relaxation of Optically Pumped Rubidium Atoms in
  Solid Parahydrogen.
\newblock \emph{\bibinfo{journal}{Physical Review Letters}}
  \textbf{\bibinfo{volume}{117}} (\bibinfo{year}{2016}).

\bibitem{Kanagin25}
\bibinfo{author}{Kanagin, A.~N.} \emph{et~al.}
\newblock Impurities in cryogenic solids: a new platform for hybrid quantum
  systems (\bibinfo{year}{2025}).
\newblock \bibinfo{note}{ArXiv:2508.21651}.

\bibitem{Venables1968}
\bibinfo{author}{Venables, J.~A.}, \bibinfo{author}{Ball, D.~J.} \&
  \bibinfo{author}{Thomas, G.~J.}
\newblock An electron microscope liquid helium stage for use with accessories.
\newblock \emph{\bibinfo{journal}{Journal of Physics E: Scientific
  Instruments}} \textbf{\bibinfo{volume}{1}}, \bibinfo{pages}{121--126}
  (\bibinfo{year}{1968}).

\bibitem{Weiss1998}
\bibinfo{author}{Weiss, G.}, \bibinfo{author}{Eschenr\"{o}der, K.},
  \bibinfo{author}{Classen, J.} \& \bibinfo{author}{Hunklinger, S.}
\newblock Ultrasonic Measurements on Quench-Condensed Noble Gas Films.
\newblock \emph{\bibinfo{journal}{Journal of Low Temperature Physics}}
  \textbf{\bibinfo{volume}{111}}, \bibinfo{pages}{321--326}
  (\bibinfo{year}{1998}).

\bibitem{Metcalf2000}
\bibinfo{author}{Metcalf, T.~H.} \& \bibinfo{author}{Pohl, R.}
\newblock Annealing of quench-condensed noble gas films.
\newblock \emph{\bibinfo{journal}{Physica B: Condensed Matter}}
  \textbf{\bibinfo{volume}{284--288}}, \bibinfo{pages}{381--382}
  (\bibinfo{year}{2000}).

\bibitem{triplePointWetting}
\bibinfo{author}{Migone, A.~D.}, \bibinfo{author}{Dash, J.~G.},
  \bibinfo{author}{Schick, M.} \& \bibinfo{author}{Vilches, O.~E.}
\newblock Triple-point wetting of neon films.
\newblock \emph{\bibinfo{journal}{Phys. Rev. B}} \textbf{\bibinfo{volume}{34}},
  \bibinfo{pages}{6322--6325} (\bibinfo{year}{1986}).

\bibitem{LeidererReview}
\bibinfo{author}{Leiderer, P.}
\newblock Surface Electrons on Solid Quantum Substrates: A Brief Review.
\newblock \emph{\bibinfo{journal}{Journal of Low Temperature Physics}}
  (\bibinfo{year}{2025}).

\bibitem{Cassidy_Ne_growth}
\bibinfo{author}{Matkovic, K.} \emph{et~al.}
\newblock Characterizing Neon Thin Film Growth with an NbTiN Superconducting
  Resonator Array (\bibinfo{year}{2025}).
\newblock \eprint{2510.21029}.

\bibitem{Pop_hydrogen}
\bibinfo{author}{Valenti, F.} \emph{et~al.}
\newblock Hydrogen crystals reduce dissipation in superconducting resonators.
\newblock \emph{\bibinfo{journal}{Phys. Rev. B}}
  \textbf{\bibinfo{volume}{109}}, \bibinfo{pages}{054503}
  (\bibinfo{year}{2024}).

\bibitem{YBCO_res1}
\bibinfo{author}{Ghirri, A.} \emph{et~al.}
\newblock YBa2Cu3O7 microwave resonators for strong collective coupling with
  spin ensembles.
\newblock \emph{\bibinfo{journal}{Applied Physics Letters}}
  \textbf{\bibinfo{volume}{106}}, \bibinfo{pages}{184101}
  (\bibinfo{year}{2015}).

\bibitem{YBCO_res2}
\bibinfo{author}{Velluire-Pellat, Z.} \emph{et~al.}
\newblock Hybrid quantum systems with high-T$_c$ superconducting resonators.
\newblock \emph{\bibinfo{journal}{Scientific Reports}}
  \textbf{\bibinfo{volume}{13}}, \bibinfo{pages}{14366} (\bibinfo{year}{2023}).

\bibitem{YBCO_res3}
\bibinfo{author}{Ghirri, A.}, \bibinfo{author}{Cavani, M.},
  \bibinfo{author}{Bonizzoni, C.} \& \bibinfo{author}{Affronte, M.}
\newblock Coherent coupling between YBCO superconducting resonators and
  sub-micrometer-thick YIG films (\bibinfo{year}{2025}).
\newblock \eprint{2506.22240}.

\bibitem{YBCO_res4}
\bibinfo{author}{Li-Bin, S.} \emph{et~al.}
\newblock A study on the microwave responses of YBCO and TBCCO thin films by
  coplanar resonator technique.
\newblock \emph{\bibinfo{journal}{Chinese Physics}}
  \textbf{\bibinfo{volume}{16}}, \bibinfo{pages}{3036} (\bibinfo{year}{2007}).

\bibitem{YBCO_discovery}
\bibinfo{author}{Wu, M.~K.} \emph{et~al.}
\newblock Superconductivity at 93 K in a new mixed-phase Y-Ba-Cu-O compound
  system at ambient pressure.
\newblock \emph{\bibinfo{journal}{Phys. Rev. Lett.}}
  \textbf{\bibinfo{volume}{58}}, \bibinfo{pages}{908--910}
  (\bibinfo{year}{1987}).

\bibitem{Ne_TP_value}
\bibinfo{author}{Stull, D.~R.}
\newblock Vapor Pressure of Pure Substances. Organic and Inorganic Compounds.
\newblock \emph{\bibinfo{journal}{Industrial \& Engineering Chemistry}}
  \textbf{\bibinfo{volume}{39}}, \bibinfo{pages}{517--550}
  (\bibinfo{year}{1947}).
\newblock \eprint{https://doi.org/10.1021/ie50448a022}.

\bibitem{trueYBCO_epi}
\bibinfo{author}{Denhoff, M.~W.} \& \bibinfo{author}{McCaffrey, J.~P.}
\newblock Epitaxial Y1Ba2Cu3O7 thin films on CeO2 buffer layers on sapphire
  substrates.
\newblock \emph{\bibinfo{journal}{Journal of Applied Physics}}
  \textbf{\bibinfo{volume}{70}}, \bibinfo{pages}{3986--3988}
  (\bibinfo{year}{1991}).

\bibitem{CeO2_buffer}
\bibinfo{author}{Wu, X.~D.} \emph{et~al.}
\newblock Epitaxial CeO2 films as buffer layers for high-temperature
  superconducting thin films.
\newblock \emph{\bibinfo{journal}{Applied Physics Letters}}
  \textbf{\bibinfo{volume}{58}}, \bibinfo{pages}{2165--2167}
  (\bibinfo{year}{1991}).

\bibitem{filmnote}
\bibinfo{note}{The films are provided by Ceraco, and StarCryo performed the
  fabrication procedures.}

\bibitem{CPW_cQED}
\bibinfo{author}{G\"oppl, M.} \emph{et~al.}
\newblock Coplanar waveguide resonators for circuit quantum electrodynamics.
\newblock \emph{\bibinfo{journal}{Journal of Applied Physics}}
  \textbf{\bibinfo{volume}{104}}, \bibinfo{pages}{113904}
  (\bibinfo{year}{2008}).

\bibitem{McRae_review}
\bibinfo{author}{McRae, C. R.~H.} \emph{et~al.}
\newblock Materials loss measurements using superconducting microwave
  resonators.
\newblock \emph{\bibinfo{journal}{Review of Scientific Instruments}}
  \textbf{\bibinfo{volume}{91}}, \bibinfo{pages}{091101}
  (\bibinfo{year}{2020}).

\bibitem{YBCO_lambda}
\bibinfo{author}{Hardy, W.~N.}, \bibinfo{author}{Bonn, D.~A.},
  \bibinfo{author}{Morgan, D.~C.}, \bibinfo{author}{Liang, R.} \&
  \bibinfo{author}{Zhang, K.}
\newblock Precision measurements of the temperature dependence of
  \ensuremath{\lambda} in
  ${\mathrm{YBa}}_{2}$${\mathrm{Cu}}_{3}$${\mathrm{O}}_{6.95}$: Strong evidence
  for nodes in the gap function.
\newblock \emph{\bibinfo{journal}{Phys. Rev. Lett.}}
  \textbf{\bibinfo{volume}{70}}, \bibinfo{pages}{3999--4002}
  (\bibinfo{year}{1993}).

\bibitem{Hirschfeld_YBCO_1}
\bibinfo{author}{Hirschfeld, P.~J.}, \bibinfo{author}{Putikka, W.~O.} \&
  \bibinfo{author}{Scalapino, D.~J.}
\newblock Microwave conductivity of d-wave superconductors.
\newblock \emph{\bibinfo{journal}{Phys. Rev. Lett.}}
  \textbf{\bibinfo{volume}{71}}, \bibinfo{pages}{3705--3708}
  (\bibinfo{year}{1993}).

\bibitem{Hirschfeld_YBCO_2}
\bibinfo{author}{Hirschfeld, P.~J.}, \bibinfo{author}{Putikka, W.~O.} \&
  \bibinfo{author}{Scalapino, D.~J.}
\newblock d-wave model for microwave response of
  high-${\mathit{T}}_{\mathit{c}}$ superconductors.
\newblock \emph{\bibinfo{journal}{Phys. Rev. B}} \textbf{\bibinfo{volume}{50}},
  \bibinfo{pages}{10250--10264} (\bibinfo{year}{1994}).

\bibitem{TLS_original}
\bibinfo{author}{Phillips, W.~A.}
\newblock Two-level states in glasses.
\newblock \emph{\bibinfo{journal}{Reports on Progress in Physics}}
  \textbf{\bibinfo{volume}{50}}, \bibinfo{pages}{1657} (\bibinfo{year}{1987}).

\bibitem{TLS_freq_shift}
\bibinfo{author}{Barends, R.} \emph{et~al.}
\newblock Contribution of dielectrics to frequency and noise of NbTiN
  superconducting resonators.
\newblock \emph{\bibinfo{journal}{Applied Physics Letters}}
  \textbf{\bibinfo{volume}{92}}, \bibinfo{pages}{223502}
  (\bibinfo{year}{2008}).

\bibitem{muller_TLS}
\bibinfo{author}{M\"uller, C.}, \bibinfo{author}{Cole, J.~H.} \&
  \bibinfo{author}{Lisenfeld, J.}
\newblock Towards understanding two-level-systems in amorphous solids: insights
  from quantum circuits.
\newblock \emph{\bibinfo{journal}{Reports on Progress in Physics}}
  \textbf{\bibinfo{volume}{82}}, \bibinfo{pages}{124501}
  (\bibinfo{year}{2019}).

\bibitem{GeYang}
\bibinfo{author}{Yang, G.} \emph{et~al.}
\newblock Coupling an Ensemble of Electrons on Superfluid Helium to a
  Superconducting Circuit.
\newblock \emph{\bibinfo{journal}{Phys. Rev. X}} \textbf{\bibinfo{volume}{6}},
  \bibinfo{pages}{011031} (\bibinfo{year}{2016}).

\bibitem{Koolstra}
\bibinfo{author}{Koolstra, G.}, \bibinfo{author}{Yang, G.} \&
  \bibinfo{author}{Schuster, D.~I.}
\newblock Coupling a single electron on superfluid helium to a superconducting
  resonator.
\newblock \emph{\bibinfo{journal}{Nature Communications}}
  \textbf{\bibinfo{volume}{10}}, \bibinfo{pages}{5323} (\bibinfo{year}{2019}).

\bibitem{Note1}
\bibinfo{note}{We attribute the difference in solidification temperature from
  the nominal triple temperature (24.56 K) to a slight miscalibration of the
  thermometer.}

\bibitem{JBA}
\bibinfo{author}{Siddiqi, I.} \emph{et~al.}
\newblock RF-Driven Josephson Bifurcation Amplifier for Quantum Measurement.
\newblock \emph{\bibinfo{journal}{Phys. Rev. Lett.}}
  \textbf{\bibinfo{volume}{93}}, \bibinfo{pages}{207002}
  (\bibinfo{year}{2004}).

\end{thebibliography}

\end{document}